# Reconciling Human Development and Giant Panda Protection Goals: Cost-efficiency Evaluation of Farmland Reverting and Energy Substitution Programs in Wolong National Reserve

Keyi Liu[a], Yufeng Chen [a,b], Liyan Xu[a]*, Xiao Zhang[a], Zilin Wang[b], Hailong Li[c]*, Yansheng Yang[d], Hong You[e], Dihua Li[a]

a. College of Architecture and Landscape, Peking University, Beijing 100871, China
b. YANGTZE Eco-Environment Engineering Research Center, China Three Gorges Corporation, Beijing 100038, China
c. Chinese Society for Urban Studies, CSUS
d. School of Tourism and Urban-Rural Planning, Zhejiang Gongshang University
e. Beijing Municipal Institute of City Planning & Design

\* Co-corresponding authors

# Abstract

Balancing human development with conservation necessitates ecological policies that optimize outcomes within limited budgets, highlighting the importance of cost-efficiency and local impact analysis. This study employs the Socio-Econ-Ecosystem Multipurpose Simulator (SEEMS), an Agent-Based Model (ABM) designed for simulating small-scale Coupled Human and Nature Systems (CHANS), to evaluate the cost-efficiency of two major ecology conservation programs: Grain-to-Green (G2G) and Firewood-to-Electricity (F2E). Focusing on China's Wolong National Reserve, a worldwide hot spot for flagship species conservation, the study evaluates the direct benefits of these programs, including reverted farmland area and firewood consumption, along with their combined indirect benefits on habitat quality, carbon emissions, and gross economic benefits. The findings are as follows: (1) The G2G program achieves optimal financial efficiency at approximately 500 CNY/Mu, with diminishing returns observed beyond 1000 CNY/Mu; (2) For the F2E program, the most fiscally cost-efficient option arises when the subsidized electricity price is at 0.4-0.5 CNY/kWh, while further reductions of the prices to below 0.1 CNY/kWh result in a diminishing cost-benefit ratio; (3) Comprehensive cost-efficiency analysis reveals no significant link between financial burden and carbon emissions, but a positive correlation with habitat quality and an inverted U-shaped relationship with total economic income; (4) Pareto analysis identifies 18 optimal dual-policy combinations for balancing carbon footprint, habitat quality, and gross economic benefits; (5) Posterior Pareto optimization further refines the selection of a specific policy scheme for a given realistic scenario. The analytical framework of this paper helps policymakers design economically viable and environmentally sustainable policies, addressing global conservation challenges.

# Keywords

# Highlights

1 Wolong Nature Reserve, exemplifies balancing development with conservation.
2 The G2G and F2E programs require aligning economic and ecological goals.
3 Employ SEEMS to predict the impacts of policies on carbon, habitat, and economics.
4 Cost-efficiency analysis finds optimal subsidies for G2G and F2E programs.
5 Overall Pareto optimization decides policy schemes for realistic situations.

# 1 Introduction

Preserving ecological diversity and promoting economic growth are both major aims among the United Nation's Sustainable Development Goals (SDGs) (United Nation, 2015), while only too often the two witness conflicts. In ecologically sensitive areas, the imperative to preserve biodiversity often clashes with the need for local development (J. Kang et al., 2021). The situation is especially critical in developing countries, where the pervasive natural resource extraction-dependent economic model creates a contest for scarce resources like land, energy, and biomass, intensifying the tension between economic development and ecological conservation (Ghoddousi et al., 2022; Zeng et al., 2023). A variety of policies for harmonizing species protection and human development has been proposed in the existing literature, such as the Ecological Compensation Program, a financial transfer mechanism that mitigates the negative influences of developing activities on the natural environment (Cuperus et al., 1996; Jeník, 2002; Wunder, 2005).

Nevertheless, two crucial challenges emerge in designing such policies, namely the precision design of the program, as well as their cost-efficiency. On the one hand, as socio-ecological systems are characterized by complex interconnections and feedback loops, where single-focus policies may produce unforeseeable consequences, designing effective conservation policies requires precision in implementation to achieve ecological objectives without unintended side effects (Andres et al., 2012; Van der Ploeg & Withagen, 2015). On the other hand, ecological preservation policies are often funded through financial transfers, but in developing countries, the demand for conservation funding usually far exceeds available resources (Balmford et al., 2003). As protecting all habitats at the highest compensation levels is not financially viable, necessitating trade-offs between scenarios with potentially conflicting objectives to achieve the best outcomes at the lowest possible cost.

China's Wolong National Reserve presents a vivid example of the above-stated challenges. As the largest and most critical habitat for the giant panda in China, Wolong is globally significant for its high biodiversity and diverse landscapes, which also support other important species like the snow leopard and golden monkey. Wolong has hence been identified as one of the 25 biodiversity hotspots for conservation priorities (Myers et al., 2000), included in Global 200 Ecoregions (Olson & Dinerstein, 2003) by Conservation International (CI) and World Wildlife Fund (WWF) respectively, and also listed as a World Heritage by the UNESCO World Heritage Centre in 2006. However, aside from its global conservation significance, Wolong has been inhabited by humans for over 500 years, with approximately 5,000 residents currently residing there. Human activities, such as agriculture and firewood collection have been pillars supporting the local self-sufficient economy (Liu et al., 1999; Ouyang et al., 2001), while the increasing magnitude of such activities with the population growth in the past few decades, along with an emerging tourism sector which also indirectly depends on natural resource extraction have increasingly threatened the fragile balance between human livelihoods and natural ecosystems. The issue of reconciling human development with species conservation faced by Wolong is so pronounced that it is listed as one of the seven most discussed regions in global Land Use and Land Cover Change (LUCC) studies (Rindfuss et al., 2007), as well as one of the six most critical sites for human-nature system coupling research globally (Liu et al., 2007). Unfortunately, despite the extensive research on giant panda conservation (D. Kang, 2022), the socioeconomic development of Wolong's human residents has been relatively insufficient (An et al., 2001, 2006; Xu et al., 2006), revealing a significant gap in designing ecological preservation policies in such regions.

In this paper, with Wolong as an exemplar case, we address the two challenges through designing and implementing a complex-system simulation-based, multi-objective decision-making system, which is an expansion of the Socio-Econ-Ecosystem Multipurpose Simulator (SEEMS) (Chen et al., 2023). By design, the system effectively responds to the above-stated two challenges.

First, the design of SEEMS follows a complex-system paradigm, such as to reflect the Complex Adaptive Systems (CAS) (Lansing, 2003; Schneider & Somers, 2006) nature of the socio-ecological system in Wolong. Particularly, Wolong is regarded as a typical Coupled Human and Natural System (CHANS), where human activities and ecological processes are interlinked through complex feedback loops, which evolve over time and are subject to the influence of various internal and external shocks, and macro-level socio-ecological patterns emerge from the aggregation of micro-scale actions (Alberti et al., 2011; Liu et al., 2007; Sheppard & McMaster, 2008). All these complications render precise prediction of the outcome of any conservation policy difficult. For example, the farmland-reverting program may facilitate a boom in grain-consuming wildlife such as boars, which intensifies the human-wildlife tension and discourages local residents from signing in the program; or, for another example, subsidizing electricity use, which is intended to suppress firewood-collecting inflicted human intervention on wildlife habitats, may prove a disserve as more tourists are attracted to the habitats by the improved accommodation thanks to the convenience brought about by electricity-powered apparatus. Fortunately, compared to the more conventional top-down approach (An & López-Carr, 2012; Field et al., 2006), the intrinsic bottom-up design of SEEMS, like other Agent-based Models (ABM) (Parker et al., 2003), helps simulate the phenomenon of "emergence", i.e., complex outcomes out of relatively simple rules of interactions between the human/societal agents, wildlife, and the natural environment (Lansing, 2003; Schneider & Somers, 2006), and is therefore especially suitable for our research purpose. Furthermore, SEEMS is tailored to a small CHANS with an open economy, easy to expand with additional policy-related modules, and is also easy to implement for its minimal data requirements. Therefore, through including conservation policies as "global controllers," SEEMS allows for a systematic evaluation of their benefits, drawbacks, and trade-offs, such as to respond to the precision policy design challenge, and to enable us to answer key research questions including what are the actual ecological, economic, and social impacts of the policies, and how can one choose the most effective policy intervention strategies to optimize socio-ecological system welfare.

Second, we integrate SEEMS within a multi-objective optimization framework, to fulfill the needs for decision-making with various direct and implicit goals, constraints, and preferences. On the one hand, ecological protection policies entail both direct benefits and complex indirect impacts. For example, the farmland-reverting program directly increases reverted farmland area, and potentially releases agricultural labor. When combined with initiatives to subsidize electricity use, enhancing the appeal of tourism as an employment sector, these two programs collectively reshape the local employment landscape. Moreover, subsidized electricity initiatives not only reduce habitat disturbances for giant pandas by limiting firewood collection, but also affect carbon emissions via changes in community energy consumption patterns. Clearly, a single-benefit evaluation method falls short in addressing these trade-offs, necessitating multi-objective optimization methods for better decision support. Pareto optimization is valuable for balancing objectives, especially conflicting ones. A solution within the Pareto optimal set is such that there is no other solution that can improve at least one objective without worsening any of the others. (Deb, 2001). This also indicates that no single solution can achieve optimality across all objectives at once. On the other hand, to make the final decision, real-world constraints, particularly budget limitations, and decision-maker preferences must be incorporated into the decision-making process (Rachmawati

& Srinivasan, 2006; Williams & Kendall, 2017). Indeed, a key challenge in policy design under limited financial resources is achieving ecological protection goals at the lowest cost. Cost-efficiency analysis is essential for identifying optimal resource allocation strategies, ensuring that limited financial resources are used to maximize policy effectiveness (Athanassopoulos & Triantis, 1998). The budget constraint line serves as a practical tool in this context. By mapping out this line, decision-makers can clearly visualize the various combinations of resource allocations that are feasible within budgetary limits (Meyer & Shipley, 1970). Indifference curves, meanwhile, are vital for understanding decision-maker preferences, which show combinations of two attributes with the same utility (Knetsch, 1989; Samuelson, 1956). They can not only display the tradeoff and substitution relationship between the two attributes, but also reveal benefit distribution via multiple curves, helping decision-makers quickly identify favorable attribute combination regions. Therefore, the study combines Pareto analysis, budget constraint lines, and indifference curves to assess the overall benefits of different ecological protection policies. This integrated approach provides decision-makers with an optimal solution that screens out the range of policy combinations that best meet the preferences of the decision-maker within the budget.

The rest of this paper is organized as follows: Chapter 2 provides a background of the Wolong National Reserve, details its specific challenges, and introduces candidate ecological protection plans to evaluate. Chapter 3 outlines the modeling strategy, including model verification and validation, performance metrics, and data sources. Chapter 4 presents the simulation results, and performs multi-objective evaluation of viable policy choices under different scenarios. Finally, Chapter 5 offers conclusions and suggests directions for future research.

## 2 Research Area and Backgrounds

### 2.1 Ecological Challenges and Policy Interventions in Wolong National Reserve

Wolong is located in a remote, mountainous area. Before the establishment of the nature reserve, the residents lived largely in autarky. With the rapid growth of China's development, the local economy has also changed. Particularly since 2000, the introduction of cash crops and eco-tourism has significantly increased residents' income, and the local economy has gradually evolved into a small and open economic system mainly based on expanding agriculture and service industries.

The natural resource-dependent economic development model in Wolong has created a competition between economic development and the protection of the habitats of giant pandas and other species. The main scarce resources at stake are energy and land. Specifically, two kinds of human activities threaten the habitat: agricultural activities and firewood collection (Liu et al., 1999; Ouyang et al., 2001), both traditional components of the self-sufficient economy. Owing to the high altitude and cold climate, energy issues plague the residents in the long term. Firewood is the traditional energy source, and its collection significantly harms the giant panda habitat. Additionally, due to population pressure, local farmers have historically converted forests and bamboo groves into farmland, severely damaging the habitat (Fig. 1). Although deforestation for agricultural expansion was forbidden in the early 2000s, existing farmland already fragments the main habitats, letting alone potential enforcement issues. Furthermore, recent economic growth has expanded the scale of these activities. The first reason is that cash crop introduction has made agriculture profitable, which further stimulates the motivation for cultivation. Additionally, the rise of tourism has increased energy demand and stimulated the demand for agricultural and specialty products, leading to further agricultural expansion. In summary, the increased range and intensity of human activities inevitably encroach upon wildlife habitats. Areas of intense human activity coincide with the primary habitat of the giant panda, posing a great threat to the habitat quality.

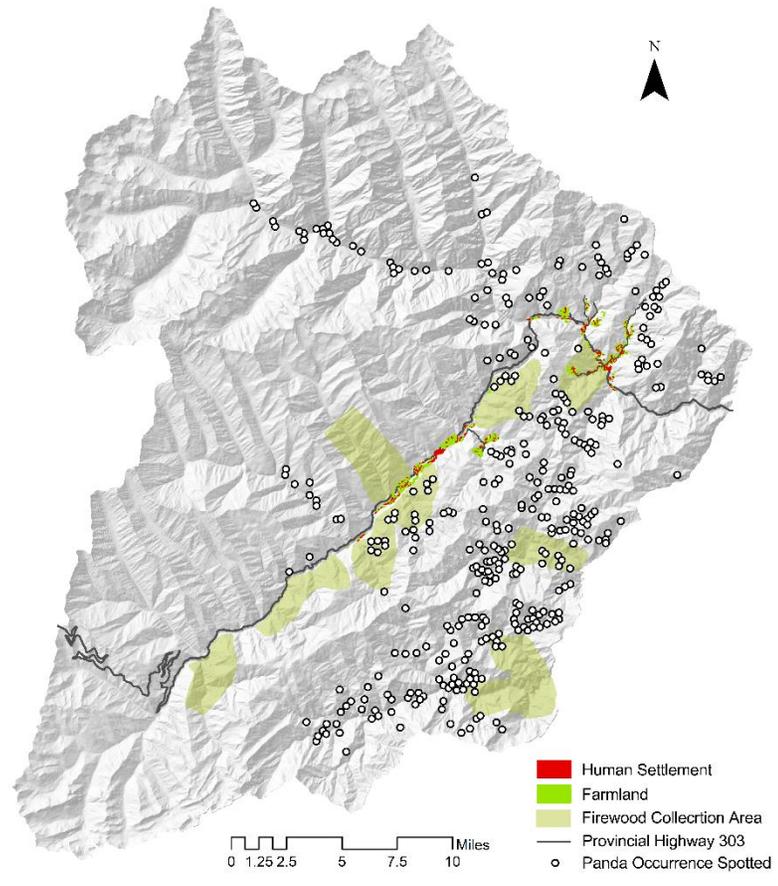

Fig. 1. The map of Wolong National Reserve. [1]

To enhance habitat conservation, the local government has implemented ecological conservation plans for energy and land use since the early 2000s. These plans encourage more sustainable and environmentally friendly production and lifestyle while compensating for potential economic losses from adopting ecological practices. Specifically, they encompass two key programs: the Grain-to-Green (G2G) program, and the Firewood-to-Electricity (F2E) program. The G2G program encourages converting farmland back to forest or bamboo groves and provides annual compensation to offset the income loss of giving agriculture up (Deng et al., 2014; Uchida et al., 2005). The F2E program involves constructing small hydropower stations to generate electricity and encouraging the use of electricity over firewood by electricity price subsidies. This plan is expected to directly reduce the local carbon footprint and indirectly aid in habitat protection by discouraging illegal firewood

---

[1] Data source: Sichuan Wolong National Natural Reserve Administration; The Third National Giant Panda Population Survey. Map lines delineate study areas and do not necessarily depict accepted national boundaries

collection (Wu et al., 2019). In addition, both programs help to unleash the labor force in rural households and offer potential new economic opportunities. For the F2E program, the energy upgrade can provide better living conditions, thus enabling farmers to enter the more lucrative homestay lodging business. In principle, both programs are voluntary, allowing households to independently decide whether to participate (Table 1).

Table 1
Details of the G2G and F2E programs

|  | G2G | | F2E | |
| --- | --- | --- | --- | --- |
|  | Participant | Non-participant | Participant | Non-participant |
| Household direct economic loss | Agriculture income loss | No | Additional payment for electricity | No |
| Household Revenues | Policy Compensation | No | Subsidized electricity price | No |
| Impact on the workforce | + | - | + | - |
| Impact on other economic opportunities | No | No | Possibility to work in a high-income industry | No |
| Impact on ecology | + | - | + | - |
| Impact on carbon footprint | - | + | - | + |

While the implementation of these policies has initiated some changes, the anticipated goals have not been fully achieved due to inherent flaws in program design and budget constraints. The G2G program, for instance, has experienced a decline in public participation, resulting in overall performance that falls short of initial expectations. Conversely, the F2E program promotes increased electricity consumption, but it cannot entirely replace firewood, and occasional firewood collection continues, undermining the protective effect. Additionally, the hydropower station construction for electricity production may introduce negative environmental externalities that further complicate conservation efforts. Therefore, revising and improving these programs is urgently needed to achieve optimal conservation outcomes. Moreover, both programs provide incentives for the local household to turn to the so-called "eco-tourism", which, though, turned out not "ecological" enough. Energy modernization per se leads to increased resource consumption and waste production (Greening et al., 2000; Herring, 2006). Also, enhanced mobility through electric vehicles can extend the spatial footprint of human disturbance, and almost inevitably attract tourists deeper into wildlife habitats, affecting wildlife habitats over a larger area and for longer durations.

The conservation policies within the Wolong National Reserve have garnered significant academic attention. Early studies investigated the impact of farming and firewood collection on giant panda habitats and regional landscapes (Bearer et al., 2008), identified spatial patterns of firewood collection (He et al., 2009; Linderman et al., 2005), and employed ABM to examine dynamic changes in population characteristics and their relationship with panda habitat (An et al., 2001, 2005, 2006). These studies have laid a solid foundation for our research but also have some limitations. Primarily, these studies are nearly two decades old, and the socio-ecological system in Wolong, particularly its economic structure, has undergone significant transformation since then. Economic shifts and the surge in tourism have greatly increased the complexity of the Wolong socio-economic-ecological system, rendering previous economic models less applicable. Moreover, although there are currently no budget constraints for any program for the moment, they will eventually face a budget cap. Therefore, local authorities must judiciously allocate transferred funds to achieve optimal long-term outcomes. The intricate interplay between the two programs introduces uncertainties in cost-benefit analyses, underscoring the critical challenge of efficiently allocating limited budgets for successful conservation success.

## 2.2 Policy Options and Scenarios Design

Apparently, price, namely the compensation or subsidy standards, plays a central role in this decision-making problem. We use a policy scenario analysis to explore optimal policy selection for the Wolong National Reserve. Each plan can implement different compensation or subsidy standards, allowing for a range of condition combinations (Table 2). The primary goal is to maximize conservation effectiveness within budget constraints, measured by habitat integrity and greenhouse gas emissions while minimizing negative impacts on the livelihoods of residents.

In the current scenario, there are no budgetary constraints, with the government temporarily covering all implementation costs. Economic development is a lower priority for the Wolong National Reserve, so our assessment focuses on comparing ecological protection performance, economic opportunities for residents, and government expenditure on compensation and subsidies. This approach aims to identify general relationships and determine the optimal policy combination for various objectives, such as maximizing conservation outcomes, achieving cost-effectiveness, and enhancing economic impacts.

For the scenario simulations, we set specific subsidy amounts based on real-world conditions. The simulated compensation range for the Grain-to-Green (G2G) program is from 0 to 2000 CNY/Mu, with simulations conducted in increments of 100 CNY. For the Firewood-to-Electricity (F2E) program, we use the standard electricity price of 0.65 CNY/kWh as the upper limit, with simulations conducted in increments of 0.05 CNY. All these ranges reflect the actual compensation and subsidy levels provided by the government currently.

Table 2
Policies Scenario matrix

| | | G2G_Compensation level (0-2000 CNY/Mu )* | | | | | | |
|---|---|---|---|---|---|---|---|---|
| | | 0 | 100 | 200 | ... | 1000 | ... | 2000 |
| F2E_Subsidized electricity price (CNY/kWh) | 0.65 | GG0/ FE0.65 | GG100/ FE0.65 | GG200/ FE0.65 | | GG100/ FE0.65 | | GG2000/ FE0.65 |
| | 0.60 | GG0/ FE0.60 | GG100/ FE0.60 | GG200/ FE0.60 | | GG100/ FE0.60 | | GG2000/ FE0.60 |
| | 0.55 | GG0/ FE0.55 | GG100/ FE0.55 | GG200/ FE0.55 | | GG100/ FE0.55 | | GG2000/ FE0.55 |
| | ... | ... | ... | ... | ... | ... | ... | ... |

| | | | | | | | |
|---|---|---|---|---|---|---|---|
| 0.40 | GG0/ FE0.40 | GG100/ FE0.40 | GG200/ FE0.40 | ... | GG100/ FE0.40 | ... | GG2000/ FE0.40 |
| 0.10 | GG0/ FE0.10 | GG100/ FE0.10 | GG200/ FE0.10 | ... | GG100/ FE0.10 | ... | GG2000/ FE0.10 |
| 0.05 | GG0/ FE0.05 | GG100/ FE0.05 | GG200/ FE0.05 | | GG350/ FE0.01 | | GG2000/ FE0.05 |

\* 1 CNY (Yuan) = 0.157 Dollars

\* 1 Mu = 0.067 Hectares

# 3 Materials and Methods

## 3.1 SEEMS: The Base Model

We use the Socio-Econ-Ecosystem Multipurpose Simulator (SEEMS) as the main evaluation tool, with necessary expansions to adapt to this research's aims. SEEMS is an agent-based model designed to simulate the operation of a small-scale, agriculture-centric, open socio-econ-ecological system and generate scenarios for assessing alternative futures. It can simulate the decision-making process of individual and household agents in economic activities, along with the associated social, economic, and ecological impacts. By monitoring the performance indicators of land use, and economic and ecological aspects, researchers can evaluate policy effectiveness by visualizing diverse outcomes. Hence, by design, SEEMS is a suitable tool for studying Wolong's ecological preservation policy-making. A comprehensive description of its baseline configuration is available in prior work (Chen et al., 2023), with this section offering an overview of its principal design concepts.

SEEMS operates across four levels: individuals, households, society, and the environment. Individuals are the fundamental agents, with behaviors encompassing birth, growth, education, marriage, childbirth and migration. Households are the basic actors, and they evolve with the dynamic changes of their members and participate in most socio-economic-ecological processes. Society serves as a container for households and their relationships, as well as the stage for all socio-economic activities and policies. The environment, in turn, forms the spatial and ecological context, interacting with socio-economic elements across various dimensions.

SEEMS comprises two main subsystems: the human system and the nature system, which interact through a series of feedback mechanisms. Furthermore, policies are viewed as global variables that influence the system in multiple ways. In certain scenarios, a household's acceptance of a policy may lead to adjustments in their productive resources. For instance, accepting an energy upgrade frees a significant portion of a household's labor force from firewood collection. In other cases, policy interventions modify decision-making parameters, such as recalculating business costs for households with subsidized electricity. These policy applications act as constraints or stimuli within SEEMS, generating distinct scenario outcomes (Fig. 2).

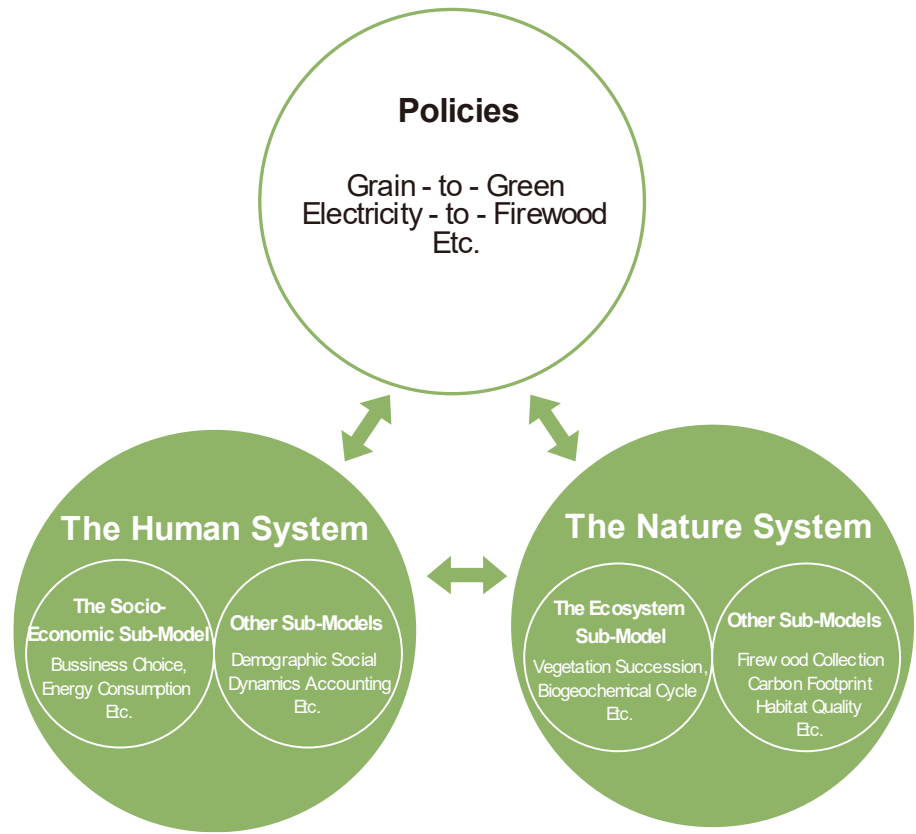

Fig. 2. The overall structure of SEEMS

Compared to other ABMs (Singh, Squire, & Strauss, 1986), the human system in SEEMS is uniquely tailored with a microeconomic foundation for small-scale and open rural economies. The fundamental concept of SEEMS emphasizes the behaviors of heterogeneous agents making decisions to maximize effectiveness, guided by their inherent preferences. The introduction of heterogeneous agents is crucial for accurately reflecting real-world situations in ABMs (Aydilek & Aydilek, 2020; Gallen, 2021). The model specifically considers two types of preferences: the income-leisure trade-off and risk attitude, which are recognized as critical determinants of production behaviors in rural economic studies (Becker, 1988; Huang, 1990). Once preferences are ensured, the household agents can take action for productive decisions. Household agents possess a set of productive resources: labor (L), land (T), and capital (K), choosing from a range of available businesses (agriculture, resource extraction, tourism, etc.) to maximize effectiveness. Households with different preferences have different optimization goals. Profit-maximizing households determine their production possibility frontier (PPF) with given L, T, and K, aiming to maximize profits within that frontier. In contrast, a leisure-maximizing household aims to minimize labor investment (L) while maintaining a baseline profit. These optimization problems can be solved through a rule-based iterative approach.

The nature system in SEEMS models the local ecosystem's functioning. Because of the inherent unpredictability of wildlife behaviors, ecological conservation objectives are assessed indirectly through simulations of habitat conditions. The main function of this system is to simulate landscape evolution in the study area, following the vegetation succession rules in the ecosystem (Clements, 1916; Curtis & McIntosh,

1951; Yunus et al., 2020).

SEEMS integrates the socio-economic-ecological system by modeling interactions between the human and nature subsystems, resulting in macro-emergent outcomes. These interactions include the impact of human activities on natural habitats (e.g., farming, housing, and infrastructure construction) and the influence of wildlife activities on human production and life. These interactions are bidirectional and iterative, capturing non-linear outcomes from a series of repetitive interactions, thereby fully reflecting the complex nature of CHANS.

### 3.2 Expansion of SEEMS

In this research, we have tailor-extended the baseline model to better accommodate the unique conditions of the Wolong National Reserve and evaluate specific policies. This section elaborates on these extensions.

#### 3.2.1 Energy Demand

One advantage of SEEMS is its ability to seamlessly integrate specialized environmental and ecological impact models as submodules. Given the focus on the F2E program, we introduced a model for calculating residents' energy demand and carbon footprint, supporting a comprehensive evaluation of socioeconomic and ecological impacts. Drawing from previous research (Zhu, 2004) and fieldwork experience, we observe that households involved in multiple businesses consume more energy both in productive activities and daily lives. Therefore, we use the number of business engagements for households as a proxy variable to calculate their energy demand. In addition, to account for potential additional energy demand generated by the "homestay lodging" industry, we introduced the factor of households' available properties for hosting guests. Considering these factors, we defined total energy demand (TEND) as the dependent variable, with the household type (determined by the number of industries: 1 industry = 1; 2 industries = 2; 3 or more industries = 3), real estate area, and the number of rooms as independent variables. A household energy demand model was fitted using field survey data from 239 households in Wolong, with the results presented in Table 3. The model exhibits a goodness-of-fit of 0.46, and all independent variables have passed the significance test at the $p < 0.1$ level.

Table 3
Household total energy demand model.

| Model | | Unstandardized Coefficients | | Standardized Coefficients | t | Sig. |
| --- | --- | --- | --- | --- | --- | --- |
| | | B | Std | β | | |
| 1 | Constant | 6.069 | .140 | | 43.451 | .000 |
| | Household type | .205 | .049 | .337 | 4.215 | .000 |
| | Area of room | .050 | .029 | .138 | 1.723 | .087 |
| | Number of rooms | .009 | .005 | .134 | 1.660 | .099 |

\* Dependent variable: Lg TEND

Following the calculation of total energy demand, we next determine the proportion supplied by

electricity—a critical factor due to the high price elasticity of electricity demand (Jorgenson, Slesnick, Stoker, & Moroney, 1987). We employ the previously defined household type proxy variable, coupled with the binary indicator for "homestay lodging industry involvement" (1 for involvement, 0 for no involvement), to calculate the household's total electricity demand (TELD). The results presented in Table 4 reveal a model with a goodness-of-fit of 0.50, with all independent variables significant at the $p < 0.1$ level. The remainder of the energy demand, unmet by electricity, is fulfilled by firewood, which, based on local survey data, has an energy equivalence of 1 kg to 2.25 kWh of electricity.

Table 4
Household total electricity demand model.

| Model | | Unstandardized Coefficients | | Standardized Coefficients | t | Sig. |
|---|---|---|---|---|---|---|
| | | B | Std | β | | |
| 1 | Constant | 5.684 | .146 | | 39.035 | .000 |
| | Area of room | .072 | .028 | .164 | 2.608 | .010 |
| | Business type | .447 | .120 | .243 | 3.716 | .000 |
| | Household type | .216 | .051 | .285 | 4.269 | .000 |

* Dependent variable: Lg TELD

### 3.2.2 Firewood Collection

To evaluate the ecological impact of household firewood demand, we explicitly model firewood collection behavior. This model not only assesses the direct effects of firewood collection on habitat but also integrates firewood collection into the economic sub-module, revealing complex dynamics within the socio-ecological system through potential chain reactions.

Firewood remains a crucial energy source for cooking and heating in rural areas of many developing countries (An, Lupi, Liu, Linderman, & Huang, 2002; Chomitz & Griffiths, 2001), and its use persists in protected areas (Liu et al., 2003). In Wolong, recent surveys indicate that following the ban on logging, residents have resorted to gathering naturally fallen wood and cutting shrubs for firewood. To simulate these real-world conditions, the model defines a grid map representing firewood resources, which have been considered constant for a few years. Villagers collect firewood in groups, and studies suggest that a mixed forest within a 90 x 90 m grid cell can sustain a household's firewood needs for approximately four years (An et al., 2005).

The firewood collection activity is modeled in two steps. The first step involves a path search algorithm that determines the route to the collection area based on cost distance, which is primarily influenced by slope resistance. The modeled area is partitioned into a grid system, and before each iteration, a partition statistics tool calculates the average resource values for each grid block. Blocks with resource values exceeding a predefined threshold are designated as firewood collection zones. Villagers aim to minimize the time spent collecting firewood by utilizing local knowledge to identify these zones. Thus, the path search behavior is characterized by local optimization with an element of randomness and is implemented using a roulette wheel algorithm. Starting from the villagers' initial location, the algorithm calculates the travel costs to adjacent grid cells and iterates until a designated firewood collection zone is reached.

The second step is a random walk-based collection behavior after reaching the area. The firewood collection behavior is set as a random walk algorithm with a step size of 1 grid until the required amount of firewood is collected. Once firewood collection is complete, the villagers return to the village, the module stops iterating, and the area of human firewood collection activities is updated (Fig. 3).

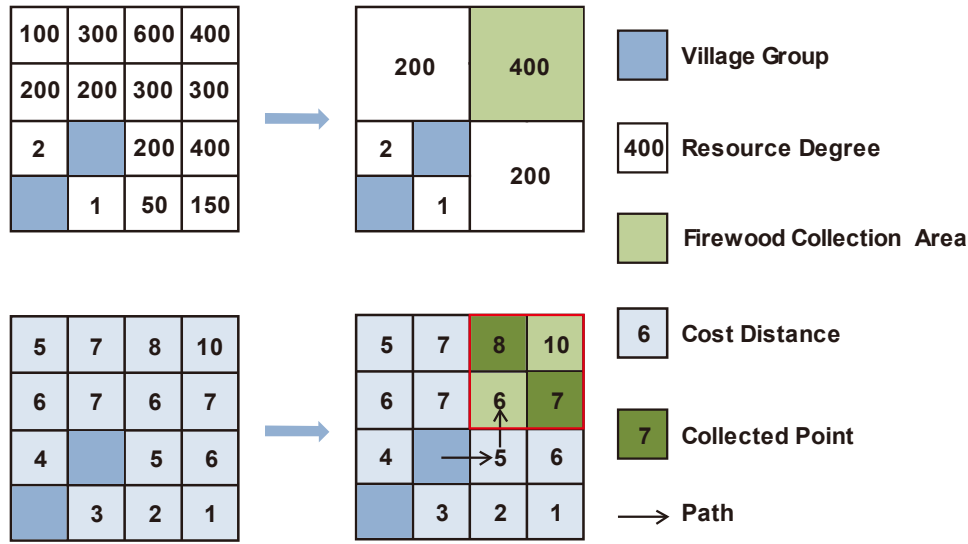

Fig. 3. Schematic Diagram of the Firewood Collection Algorithm

## 3.3 Comprehensive Policy Evaluation

### 3.3.1 Evaluation Indicators

The model generates a range of statistical data to evaluate its operational performance. In the single-policy evaluation of G2G and F2E programs, the focus is placed on policy expenditures and their direct effects, which are reflected by reverted farmland area and firewood consumption respectively. For the dual-policy evaluation, which examines the combined impacts of these two programs, the analysis extends to their comprehensive impacts on the Wolong socio-ecological system. These impacts are assessed using three key indicators: carbon footprint, habitat quality, and gross economic benefits. Among all the above indicators, habitat quality is derived from agricultural and firewood collection activities and the extent of vegetation succession, while others are directly based on the model outputs.

(1) Carbon Footprint: Based on the carbon calculator provided by the Ministry of Science and Technology of the People's Republic of China[2], we adopt carbon dioxide emission factors of 1.4375 kg $CO_2$/kg for firewood and 0.96 kg $CO_2$/kWh for electricity, to calculate the local carbon footprint. The carbon calculator, published in 2008, provides an accurate depiction of the national average carbon emission levels during that period.

---

[2] The carbon calculator provided by the Ministry of Science and Technology of the People's Republic of China http://www.acca21.org.cn/eser/counter/index.htm

(2) Habitat Quality: This study employs the methodology of Li et al. (2010), which takes into account the species' habitat preferences, including preferences for flat or gently sloping areas with bamboo and forest cover, and the influence of human activities like farming, firewood collecting, and transportation on giant panda's behavior. In the assessment indicator system, seven key factors were identified from topographical, biotic, and anthropogenic disturbance perspectives, including slope, proximity to streams, land cover types, bamboo species, and distances to roads, farmlands, and residential areas. After determining the weights of each indicator using the Analytic Hierarchy Process (AHP), a weighted composite analysis in ArcGIS was conducted with variable maps to produce an integrated habitat quality map. The overall habitat quality index for the study area was derived by aggregating the quality values across all grid cells.

(3) Gross Economic Benefits: This indicator is derived by subtracting "Financial Burden" from "Gross Economic Revenues", both generated directly by the model. "Financial Burden" refers to the aggregate expenditure of both programs and "Gross Economic Revenues" denotes the collective income of all households across various business sectors, mainly including agriculture, temp job, and lodging. The indicator "Gross Economic Benefits" focuses on comprehensively reflecting the expenditure and the corresponding economic influence, instead of the real economic benefits of policy implementation.

### 3.3.2 Evaluation Methods

This article delves into two methodologies—Cost-efficiency analysis and Multi-Objective Optimization—that are meticulously applied to evaluate the performance of different policy initiatives. By understanding how these methods are utilized, we can gain valuable insights into the optimal allocation of resources and the achievement of policy objectives.

(1) Cost-efficiency analysis: Essential in both public administration and the private sector, this evaluative tool serves as a critical gauge of an organization's, system's, or service's resource utilization efficiency in achieving defined outcomes (Piacenza, 2006). It focuses on the interplay between inputs and outputs, aiming to optimize output at minimal cost or to enhance output under input constraints. In our study, policy financial burdens are categorized as inputs, with outputs tailored to the specific programs analyzed. For example, in the G2G program, the output is measured by the area of reverted farmland, while in the F2E program, it is measured by electricity consumption. In dual-policy fiscal analysis, outputs include the three core objective indicators outlined in section 3.3.1. By aligning output metrics with program characteristics, we ensure a more accurate and comprehensive assessment of policies, avoiding financial waste.

(2) Multi-Objective Optimization
  a) Pareto analysis: Recognized as a critical approach for Multi-Objective Optimization (MOO), this method has garnered significant attention for its efficacy in identifying trade-off solutions across various objectives. Central to Pareto analysis is the concept of Pareto dominance (Gunantara, 2018), where a solution is said to dominate others if it is not inferior in all objectives and superior in at least one. This comparative analysis across objectives enables the identification of a set of Pareto optimal solutions, constituting the Pareto frontier, which provides insights into the trade-offs among objectives. In our study, we employed Pareto analysis to assess the collective impact of the G2G and F2E programs on Wolong and to explore optimal policy combination scenarios across the three objectives outlined in section 3.3.1. This approach facilitates a nuanced understanding of the trade-offs and synergies between the policies.
b) Posterior Pareto optimization: To make a choice based on the trade-offs observed in the Pareto frontier set, this study uses target lines, budget control lines, and indifference curves to assess policy benefits holistically. The target and budget control lines constrain direct effects like reverted farmland area and firewood consumption. The target line sets minimum policy requirements, such as the required minimum reverted land area for G2G programs. The budget control line reflects possible direct benefit combinations within a specific budget. In this study, total budget expenditure equals the sum of the G2G and F2E project budgets. The G2G budget is calculated by multiplying the compensation per unit area by the area of returned farmland, while the F2E budget is calculated by multiplying subsidies per kWh by electricity consumption. Due to the complex nonlinear relationship between firewood consumption and F2E policy expenditure, this overall optimization analysis uses electricity consumption as a direct effect indicator of the F2E program instead of firewood consumption. The indifference curve primarily captures the indirect effects of policy, drawn by establishing the functional relationship between the direct benefits and the three indirect influencing effects discussed in section 3.3.1.

## 3.4  Data

SEEMS relies on accurate data describing the behavioral characteristics of agents and general socio-ecological conditions in the study area. Behavioral data were derived from roughly 1000 hours of field surveys and 239 household interviews, representing about one-fifth of the households in Wolong. Qualitative and quantitative data were collected via structured interviews and questionnaires. Furthermore, data on topography, land use/land cover, vegetation, demographics, and socio-economic statistics were provided by the National Nature Reserve Administration and local authorities in Wolong. The data were organized into four datasets: individuals, households, industries, and land use. For detailed descriptions of the data structure and content, see the Supplementary Materials (Table S1).

## 3.5  Simulation, Indicator Output, Validation, and Uncertainty Analysis

After the 2008 earthquake, significant destruction occurred to local farmland, housing, and other capital assets. These transformations led to a significant shift in Wolong's socio-economic-ecological system in 2008. Therefore, we selected 2010, the year marking the preliminary completion of post-disaster reconstruction, as the starting point for our simulation. This choice also allows a long enough time range for retrospective model validation. The simulation spans a 14-year period, which is suitable for capturing medium to short-term variability in model parameters, minimizing the influence of long-term changes in parameters, such as economic income, and aligning with the vegetation succession cycle.

It is important to note that the study employs a stochastic algorithm, introducing uncertainty in conclusions at the micro-agent level, such as specific land use changes, family incomes, savings, debt status, and electricity usage. Therefore, conclusions are statistically valid only at the societal level. To minimize uncertainty from randomness, the model was run 30 times, and the average outcomes were considered the definitive results.

Detailed output indices are provided in the Supplementary Materials (Table S2).

A primary method for validating an agent-based model (ABM) is to infer and examine the accuracy of its predictions. For model validation, we use comprehensive demographic and economic data from the Wolong region, grounded in empirical observations. Calibration against historical data is typically the most effective approach, where a past reference point is selected to initiate model runs and assess how well the outcomes align with actual data. Specifically, our model validation employs metrics such as population and household growth rates, as well as economic growth rate and structure. The validation results for the baseline model, as detailed in the SEEMS publication (Chen et al., 2023), demonstrate broad agreement with observed real-world scenarios. Since the extensions introduced in this study do not significantly alter the baseline model's core socio-economic and ecological dynamics, the baseline model's validation supports the reliability of the current model.

Given the complexity of ABMs, comprehensive validation using all output indicators is generally impractical. In this study's expanded model, parameters such as firewood-to-electricity conversion rates and carbon footprint equivalents are chosen based on empirical experience. While these parameters affect the absolute values of simulation outcomes, our primary focus is on the relative performances and the sensitivity of outcomes to changes in inputs and policy conditions. Recognizing that validation poses inherent challenges for all ABMs, researchers often rely on "common sense" as a last resort, which is also adopted in this study (Brown, Page, Riolo, Zellner, & Rand, 2005; Robinson, Brown, & Currie, 2009).

# 4  Results

## 4.1  Cost-Efficiency Analysis: The Grain-to-Green (G2G) Program

To prevent potential interference from the F2E policy on the simulation outcomes of the G2G program, this study set the subsidized electricity price level at 0.65 CNY/kWh, effectively simulating a scenario without electricity subsidies Detailed data are shown in the Supplementary Materials (Table S3).

We illustrate the variations in policy expenditure, land use change, carbon footprint, and habitat quality across a compensation range from 0 to 2000 CNY/Mu (Fig. 4). Regarding policy cost-effectiveness, Figs. 4a and 4b demonstrate a rising trend in both total financial cost and reverted farmland area as compensation levels increase. Specifically, upon the compensation rising from 400 to 600 CNY/Mu, there is a significant expansion in both indicators. Additionally, upon reaching the compensation of 1000 CNY/Mu, a threshold effect emerges: the area of reverted farmland continues to increase with rising compensation up to this level, after which the rate of growth declines annually. To further analyze the relationship between policy expenditure and reverted farmland area, we conducted a focused analysis of 2024 data (Fig. 4c). The findings indicate a monotonically increasing non-linear correlation between the two indicators, with a breakpoint around 600 CNY/Mu, identifying this as the relatively optimal subsidy level. Beyond this threshold, higher compensation yields diminishing marginal returns for expanding farmland reversion. The compensation may exceed the agricultural revenue, calculated based on the opportunity cost of labor, potentially leading to wasted funds.

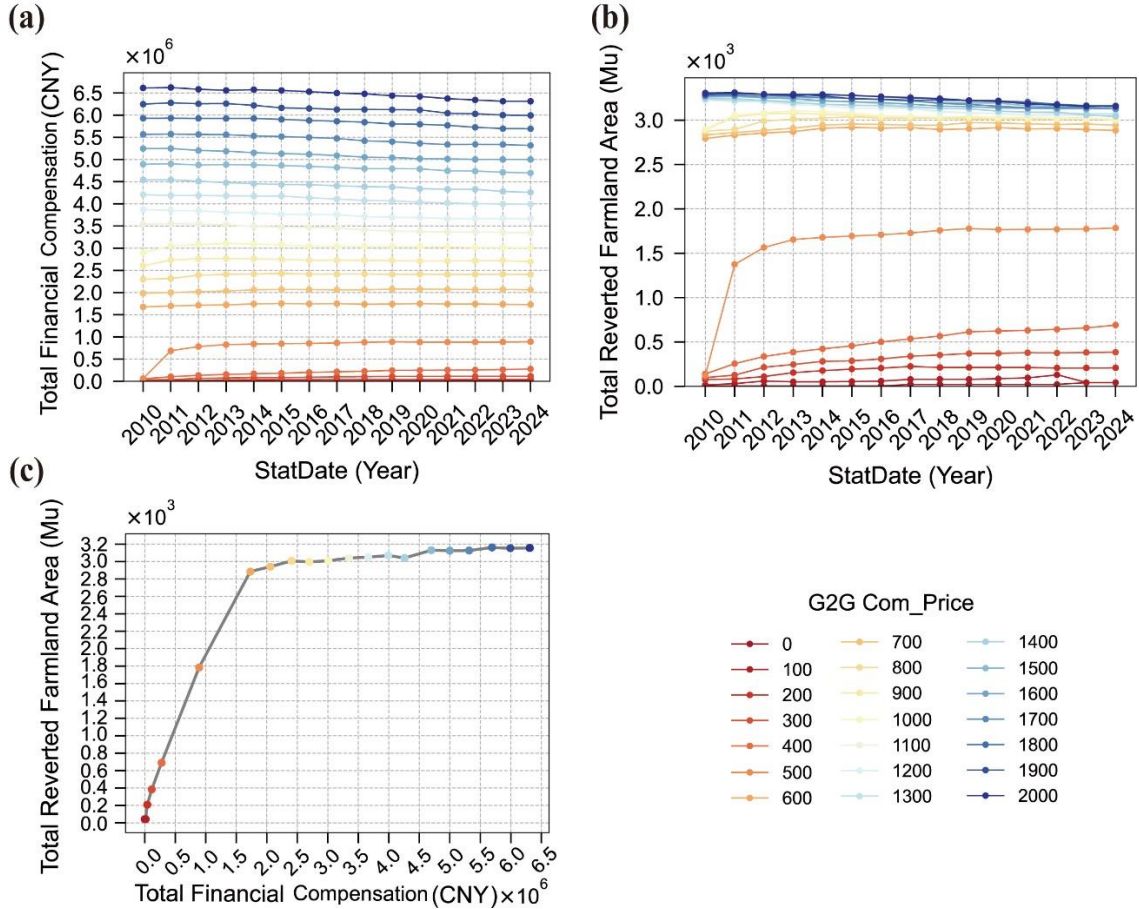

Fig. 4. The cost-efficiency analysis of the G2G program: (a) G2G Total Financial Compensation by G2G compensation Over Years, (b) Total Reverted Farmland Area by G2G compensation Over Years, (c) G2G Total Financial Compensation V.S. Total Reverted Farmland Area in 2024.

## 4.2    Cost-Efficiency Analysis: The Firewood-to-Electricity (F2E) Program

Similar to the analysis of the G2G program, we set the G2G compensation price at 0 CNY/Mu to evaluate the isolated impact of the F2E program. Detailed data are shown in the Supplementary Materials (Table S4).

Fig. 5a shows the positive correlation between total subsidies paid to households and subsidized electricity price levels, with an upward trend for F2E financial subsidy over the years. Additionally, when the subsidized electricity price drops to 0.3 CNY, a similar reduction in price results in a more substantial increase in policy expenditure. Fig. 5b reveals the complex nonlinear relationship between firewood consumption and subsidized electricity prices. Moderate price levels of 0.4 and 0.5 CNY/kWh result in lower firewood consumption, whereas higher (0.65 and 0.6 CNY/kWh) and lower (0.1 CNY/kWh) levels are associated with significantly higher firewood consumption. Although a subsidy of 0.05 CNY/kWh corresponds to the lowest firewood consumption, it also requires a substantially higher cost. Using 2024 data as an example (Fig .5c), further reducing the electricity price to 0.2 CNY/kWh or 0.05 CNY/kWh achieves even lower firewood consumption, but the costs are 4-10 times higher than those required at the 0.4-0.5 CNY/kWh level. Therefore, the optimal cost-effective subsidized electricity price range for the F2E program is between 0.4 and 0.5 CNY/kWh.

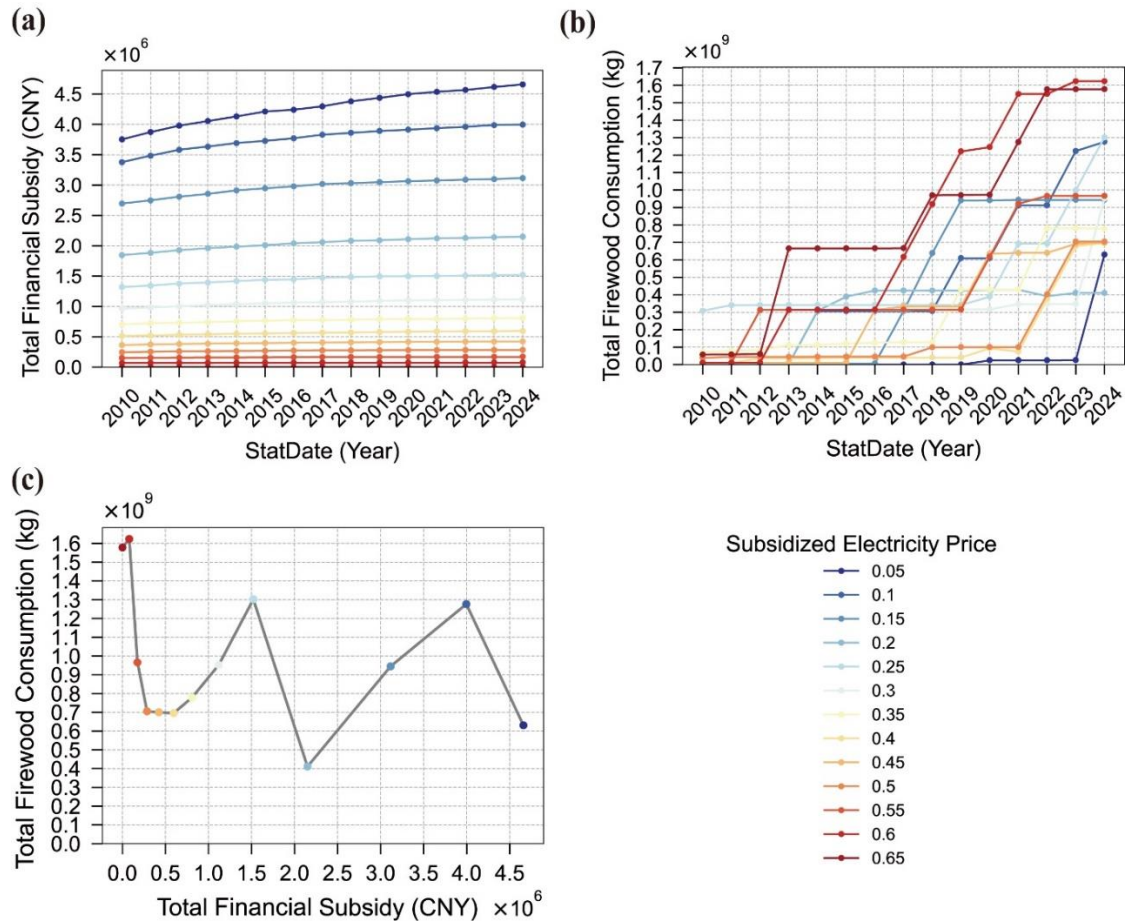

Fig. 5. The cost-efficiency analysis of the F2E program: (a) F2E Financial Subsidy by Electricity Subsidy Over Years, (b) Total Firewood Consumption by Electricity Subsidy Over Years, (c) F2E Financial Subsidy V.S. Total Firewood Consumption in 2024.

## 4.3 Dual-Policy Analysis

### 4.3.1 Comprehensive Cost-Efficiency Analysis

To analyze the cost-effectiveness under dual policy interventions for various scenarios, we used 2024 as a reference year and plotted the relationship between the financial burden of G2G and F2E programs and the three objectives of carbon footprint, habitat quality, and total financial revenue. For detailed data on outcomes, see the Supplementary Materials (Table S5).

Fig. 6 (a) illustrates the relationship between total program expenditure and carbon emissions[3]. Overall, no discernible trend is evident between total policy expenditure and carbon emissions. In the high expenditure range, however, the prevalence of blue data points indicates a marked decline in the cost-effectiveness of carbon reduction when the subsidized electricity price in the F2E program drops to 0.15 CNY/kWh or lower. Concurrently, within the low to moderate expenditure range, the dual policy impacts on carbon emissions are diffuse, with emissions likely influenced by multiple factors beyond these policies alone.

---
[3] The coordinate axes of the carbon footprint in the graph were reversed, to improve the readability of the figure.

Fig. 6 (b) demonstrates a positive correlation between total program expenditure and habitat quality. Notably, the effect of G2G subsidies on both indicators is consistent across varying F2E subsidy levels, which means under the same G2G compensation level, the level of subsidized electricity price has little impact on habitat quality. Consequently, increasing the proportion of G2G compensation under a fixed expenditure more effectively enhances habitat quality.

Fig. 6 (c) illustrates the relationship between total program expenditure and total economic income. Overall, total economic income initially declines and subsequently rises with increasing policy expenditure. At subsidized electricity prices above 0.25 CNY/kWh, subsidy levels in both programs markedly impact both total expenditure and economic income. The data distribution suggests a non-linear relationship between these indicators, aligning with the prior analysis of farmland reversion's economic impact: at lower policy expenditure levels, increased spending may reduce economic income, possibly due to insufficient subsidies to offset economic losses from decreased incentives to pursue additional income. However, when policy expenditure reaches a certain threshold, economic income starts to rise, possibly reflecting the gradual emergence of policy benefits.

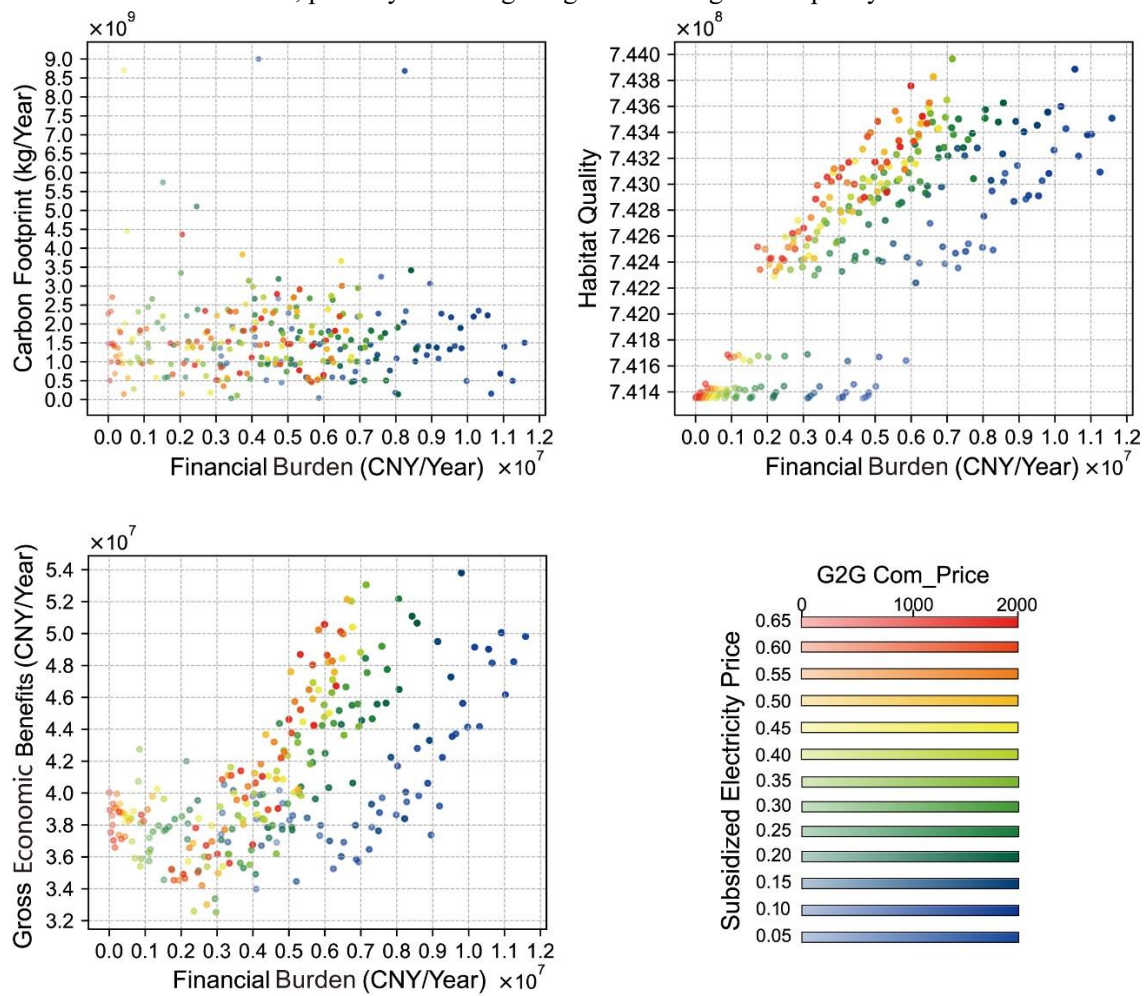

Fig. 6. The Cost-efficiency analysis of objectives: (a) Financial Burden V.S. Carbon Footprint in 2024, (b) Financial Burden V.S. Habitat Quality in 2024, (c) Financial Burden V.S. Gross Economic Benefits in 2024

### 4.3.2 Pareto Frontiers Identification

Pareto analysis is a systematic approach for evaluating and selecting optimal balances among multiple objectives. Identifying Pareto-optimal policy combinations allows for a deeper understanding of the potential impacts of varying policy choices. In this study, we used 2024 simulation data to perform a Pareto optimization analysis on three primary indicators: carbon footprint, habitat quality, and gross economic benefits. By calculating Pareto dominance relationships among policy combinations, we identify solutions that are not dominated across any of the objectives, thereby forming a Pareto frontier. This frontier represents the optimal balance achievable under current conditions. By this method, we identified approximately 18 Pareto-optimal policy combinations from the full set of policy options (Fig. 7)[4], achieving the best balance on the three considered objectives.

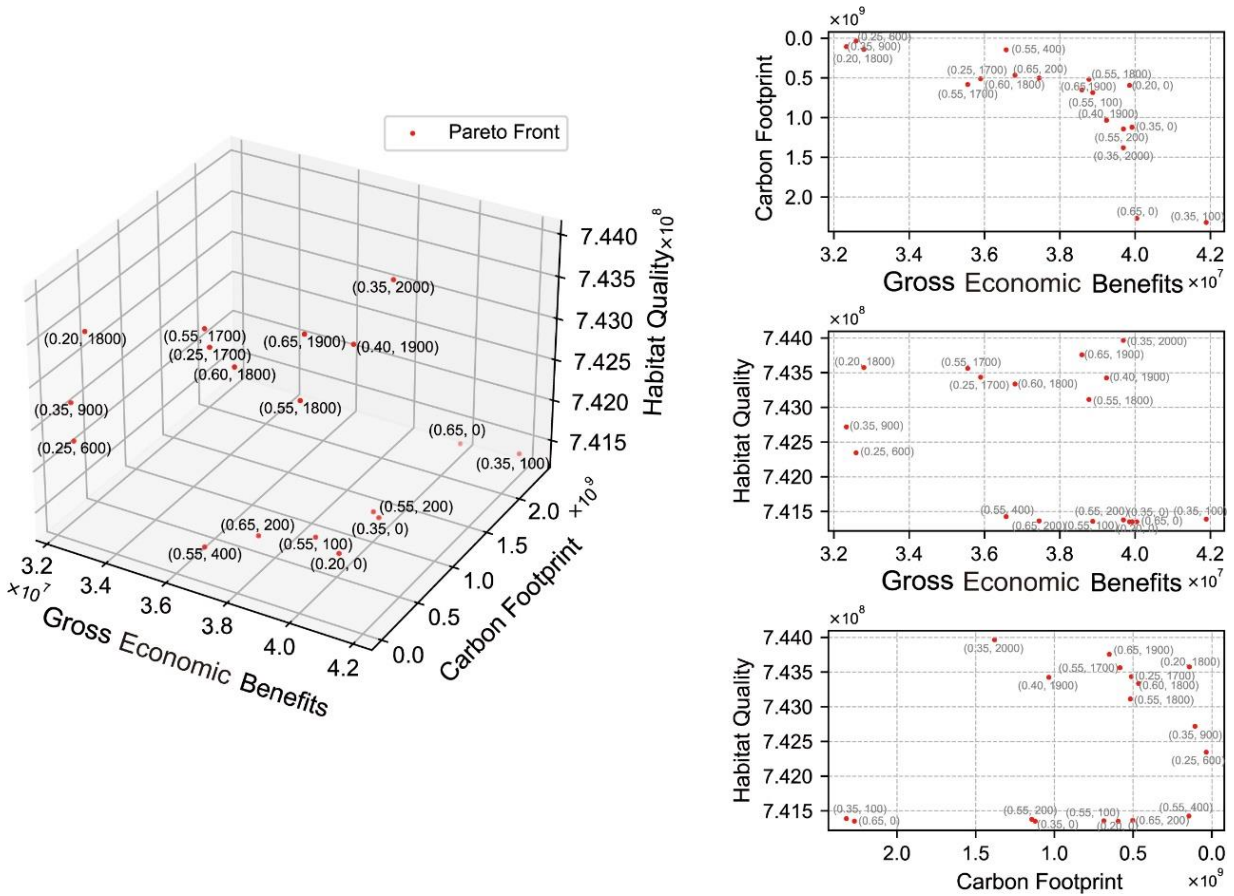

Fig. 7. Pareto Analysis of Gross Economic Benefits, Carbon Footprint, and Habitat Quality in 2024

To clearly represent the performance of each Pareto-optimal policy combination on the three target indicators, we divide the overall distribution range of the best combinations on each indicator into four equal parts: '--, -, +, ++', representing performance from relatively worst to relatively best (Table. 5).

---

[4] The coordinate axes of the carbon footprint in the graph were reversed, to improve the readability of the figure.

Table 5

Optimal policy combinations and their performances on target indicators.

|    | F2E  | G2G  | Gross Financial Benefits | Carbon Footprint | Habitat Quality |
|----|------|------|--------------------------|------------------|-----------------|
| 1  | 0.35 | 900  | --                       | ++               | +               |
| 2  | 0.25 | 600  | --                       | ++               | -               |
| 3  | 0.20 | 1800 | --                       | ++               | ++              |
| 4  | 0.55 | 400  | -                        | ++               | --              |
| 5  | 0.55 | 1700 | -                        | +                | ++              |
| 6  | 0.25 | 1700 | -                        | ++               | ++              |
| 7  | 0.60 | 1800 | -                        | ++               | +               |
| 8  | 0.65 | 2000 | +                        | ++               | --              |
| 9  | 0.55 | 1800 | +                        | ++               | +               |
| 10 | 0.65 | 1900 | +                        | +                | ++              |
| 11 | 0.55 | 100  | +                        | +                | --              |
| 12 | 0.40 | 1900 | +                        | +                | ++              |
| 13 | 0.55 | 200  | +                        | +                | --              |
| 14 | 0.35 | 2000 | +                        | -                | ++              |
| 15 | 0.20 | 0    | ++                       | +                | --              |
| 16 | 0.35 | 0    | ++                       | +                | --              |
| 17 | 0.35 | 100  | ++                       | --               | --              |
| 18 | 0.65 | 0    | ++                       | --               | --              |

### 4.3.3 Comprehensive Pareto Optimization

Classifying the performance of the Pareto solution set with respect to explicit policy goals helps decision-makers choose specific policy options based on their preferences. However, to find the overall optimal solution that considers not only the explicit policy goals but also the comprehensive socio-economic, ecological, and fiscal outcomes and constraints, additional in-depth analysis is necessary. Hence, we further offer a framework for posterior optimization based on budget constraints, policy objectives, and effect preferences. To plot the budget constraint lines and indifference curves, this study established the functional relationships between the project's direct benefits—reverted land area and electricity consumption—and the policy expenditure, along with three indirect factors in Pareto analysis. The specific functional relationships and fitting effects are detailed in Table 6. Except for the Gross Economic Benefits models, which have relatively low $R^2$ values (explaining only 50% of data variability), the other models adequately account for the data variability. For the habitat quality model, cubic fitting, as opposed to quadratic fitting, can markedly enhance the $R^2$ value from 0.8851 to 0.9192, indicating a better fit to the data. The carbon emission equivalent curve wasn't plotted because of the complex nonlinear relationship between carbon emissions and the two direct benefits and the difficulty in finding a fitting function. The carbon emission status for specific scenarios can be determined by the colors of Pareto optimal solution points.

Table 6

Model Function Expressions and Fitting Performance Metrics.

| | Model | | Train R² | Test R² |
|---|---|---|---|---|
| G2G Budget Model | $Y = a \times e^{bX_1} + c$ | $a = 1.8845$, $b = 0.0047$, $c = 226620.6$ | 0.9485 | 0.8962 |
| F2E Budget Model | $Y = aX_2^2 + bX_1 + c$ | $a = 1.9104 \times 10^{-8}$, $b = 0.5094$, $c = -824291.4$ | 0.9965 | 0.9979 |
| Habitat Quality Model | $Y = \beta_0 + \beta_1 X_1 + \beta_2 X_2 + \beta_3 X_1^2 + \beta_4 X_1 X_2 + \beta_5 X_2^2 + \beta_6 X_1^3 + \beta_7 X_1^2 X_2 + \beta_8 X_1 X_2^2 + \beta_9 X_2^3$ | $\beta_0 = -0.9578$, $\beta_1 = 1.5254$, $\beta_2 = 0.0776$, $\beta_3 = 1.5801$, $\beta_4 = -0.0047$, $\beta_5 = 0.0403$, $\beta_6 = 0.4903$, $\beta_7 = -0.0082$, $\beta_8 = 0.0200$, $\beta_9 = -0.0423$ | 0.9192 | 0.8914 |
| Economic Benefits Model | $Y = \beta_0 + \beta_1 X_1 + \beta_2 X_2 + \beta_3 X_1^2 + \beta_4 X_1 X_2 + \beta_5 X_2^2$ | $\beta_0 = -0.2455$, $\beta_1 = -0.1137$, $\beta_2 = -0.3466$, $\beta_3 = 0.3029$, $\beta_4 = 0.0281$, $\beta_5 = -0.086$ | 0.5010 | 0.5000 |

\* $X_1$: Total reverted farmland area
\* $X_2$: Total electricity consumption

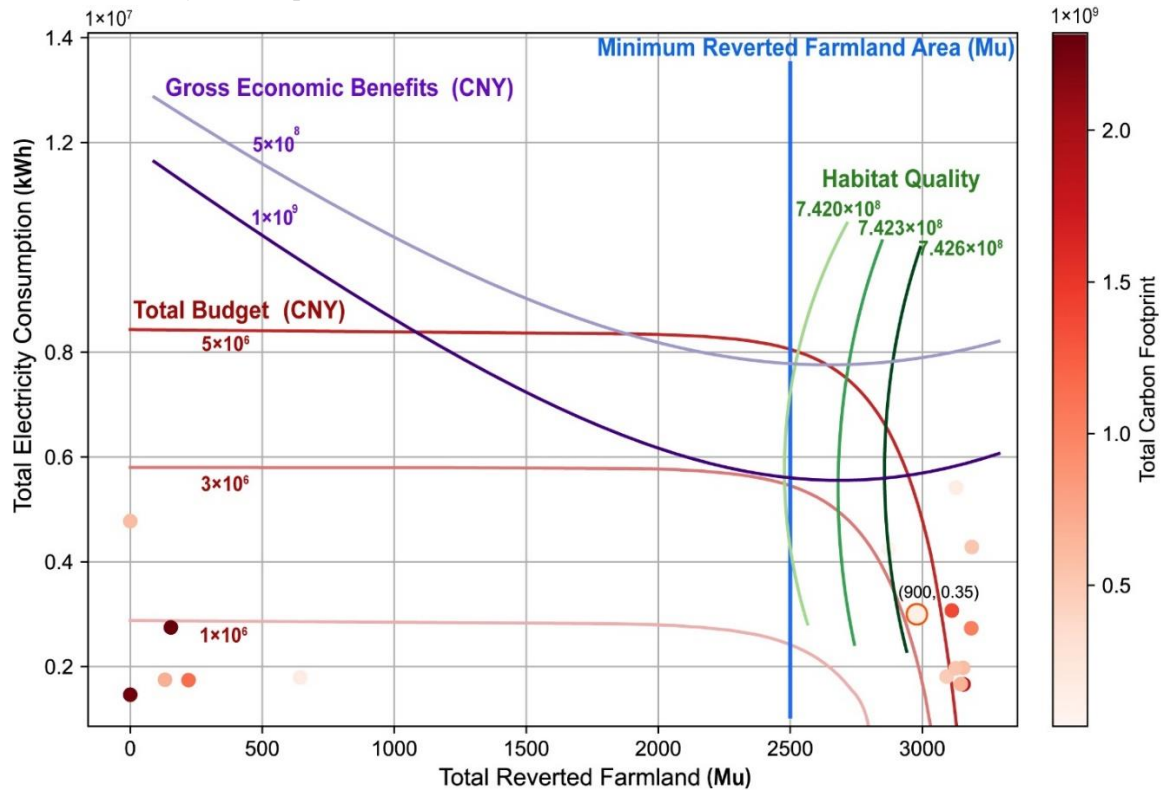

Fig. 8. Overall Pareto Optimization in 2024

As shown in Figure 8, when the real-world conditions are, for example, that the budget has to be capped at 5 million CNY, the annual reforested area is required to exceed 2,500 Mu, and greater emphasis is placed on habitat quality and carbon emissions, the policy combination with a G2G compensation of 900 CNY/Mu and a subsidized electricity price of 0.35 CNY/kWh turns out to be the one that aligns most closely with these requirements.

Figure 8 also delineates the interplay between direct and indirect benefits. The total budget curve shows when the area of reverted farmland falls below 2000 Mu, its influence on the total policy budget is negligible. The gross economic benefits curve reveals an inflection point at approximately 2700 Mu of reverted land. Surpassing this threshold necessitates an increase in both reverted land and electricity consumption to sustain equivalent economic benefits, which means further reallocating agricultural labor to other sectors is unlikely to yield substantial economic gains. The habitat quality curve shows minimal sensitivity to electricity consumption, with the minimum reverted farmland area required to maintain a given level of habitat quality occurring at around 7 million kWh of electricity consumption.

# 5 Conclusion and Discussion

This research delves into achieving a balance between ecological conservation and economic growth within the constraints of a limited budget. By developing the Socio-Econ-Ecosystem Multipurpose Simulator (SEEMS), we model the implementation outcomes of Grain-to-Green (G2G) and Firewood-to-Electricity (F2E) programs at the Wolong National Reserve.

The individual policy cost-efficiency analysis uncovers the nonlinear relationship between policy outcomes and compensation levels, with optimal effects observed at approximately 500 CNY/Mu for G2G and a subsidized electricity rate of 0.4-0.5 CNY/kWh for F2E. Nonetheless, beyond a certain threshold, the incremental benefits of increased compensation or subsidies diminish, underscoring the significance of identifying the most cost-effective policy design.

The dual-policy comprehensive cost-efficiency analysis reveals that total financial burden is not strongly correlated with carbon emissions but was positively associated with habitat quality. A noteworthy phenomenon emerges that gross economic benefits are lowest at mid-range financial expenditure levels. This phenomenon can be explained by integrating changes in economic structure with revenue fluctuations. At lower compensation thresholds, farmers may experience increased financial stress, prompting them to seek additional income streams, such as non-agricultural employment, to offset the income loss from abandoning farming and enhance overall income. At higher compensation standards, households with labor-dominant profiles are more likely to engage in the project, receive periodic compensation, and transition to other sectors, which can augment their business revenue and boost overall earnings. However, at medium subsidy standards, farmers rely on compensations to sustain a basic livelihood, diminishing their drive to seek extra income sources. Additionally, the compensation is insufficient for a complete transition to other sectors, limiting the potential growth of total income. This result highlights an intriguing principle of policy implementation: full non-enforcement or stringent enforcement is preferred. A seemingly moderate approach may lead to the most undesirable outcomes.

Of course, this is merely a reflection of a single target, and it does not justify the outright dismissal of policy options associated with moderate financial burden levels. Instead, a multi-objective assessment is required to gauge the appropriate and precise intensity of policy enforcement, which underscores the significance of Pareto analysis. Pareto analysis inherently results in a solution set with multiple non-dominated solutions, none of which are optimal across all objectives. Consequently, real-world constraints and decision-maker preferences serve as the tie-breaking points among these solutions. By considering both direct and indirect benefits comprehensively, we introduce the target constraint line, budget control line, and indifference curve. Through Posterior Pareto optimization, we can further select a policy combination that aligns more closely with actual situation constraints and objectives.

This study offers a comprehensive analysis of ecological conservation policies in the Wolong National Reserve, highlighting two main aspects of significance. First, as a critical node within the human-biosphere network, Wolong plays a vital role in global ecosystem conservation efforts. Second, the challenges faced by Wolong—balancing economic development with ecological protection in a sensitive environment—are common issues encountered throughout the developing world. These challenges often involve managing complex natural ecosystems amid growing economic demands and population pressures, while navigating the unintended consequences of narrowly focused public policies. Effective conservation requires careful trade-offs between competing programs, all within the constraints of limited budgets. In this light, Wolong serves as a model case for understanding the political-ecological dilemmas that arise in similar contexts. Addressing these challenges in Wolong could provide valuable insights for other regions worldwide facing comparable ecological and developmental tensions.

# Funding

This work was supported by National Key Research and Development Program of China (Grant No. 2022YFC3800803) and the National Natural Science Foundation of China (Grant No. 31972940).